\documentstyle[11pt]{article}
\title{\bf Axioms for Mach's mechanics}
\author{Adonai S. Sant'Anna\thanks{To whom correspondence should be sent. E-mail: adonai@scientist.com URL: http://www.geocities.com/adonaisantanna/adonaiss.html} \and Clovis A. S. Maia\thanks{E-mail: clovis@fisica.ufpr.br}}
\date{\it Dep. Matem\'atica, UFPR, C.P.
019081, Curitiba, PR, 81531-990, Brazil.}
\begin{document}
\maketitle
\newtheorem{definicao}{Definition}
\newtheorem{teorema}{Theorem}
\newtheorem{lema}{Lemma}
\newtheorem{corolario}{Corolary}
\newtheorem{proposicao}{Proposition}
\newtheorem{axioma}{Axiom}
\newtheorem{observacao}{Observation}

\begin{abstract}
We present an axiomatic framework for what we call Mach's mechanics, inspired on the ideas by A. K. T. Assis and P. Graneau about relational mechanics. We show that contrarily to what is suggested by these authors, Mach's principle does not need to be committed with any Weber's like gravitational force. Actually it is possible to settle an axiomatic framework for non-relativistic classical particle mechanics which incorporates Mach's principle and is consistent with any force that obeys the newtonian action-reaction principle. We make further criticisms concerning relational mechanics.
\end{abstract}


\section{Introduction}

	Discussions about the foundations of physics seem to be endless. From Aristotelian mechanics to superstrings, there have been always criticisms, doubts, writings, and discussions about the philosophical, physical, logical and mathematical foundations of physics. This journal itself supports such discussions. But if we take a look on the history of mechanics, which corresponds to just a portion of physics, we can see how difficult is to understand even classical mechanics, which is, maybe, the most intuitive of all physical sciences. An excellent review that illustrates these ideas is \cite{Dugas-55}. We are not only talking about the history of classical mechanics. We are talking about the great difficulty to coordinate the relations between mathematical models and physically meaningful sentences.

	Probably the best way to settle a common language in order to make such foundational discussions easier is through the correct (implicit or explicit) use of the axiomatic method. In this paper we are focused on the problem of the origin of inertia. We want to discuss about this subject from the axiomatic point of view.

	Discussions about the foundations of classical mechanics are not new. Some late 19th century physicists, for example, considered that force was an anthropomorphic or metaphysical concept that should be eliminated from mechanics. There was a strong influence of Kantianism among several german physicists like E. Mach, H. L. Helmholtz and H. R. Hertz, regarding the coordinating relations of thoughts to facts. The concept of force seemed to have a hypothetical and unobservable status. In Hertz's {\em Principles of Classical Mechanics} \cite{Hertz-56}, originaly published in 1894 \cite{Hertz-94}, the author showed to be dissatisfied with the obscurity lurking in the concept of force. He presented a formulation of classical particle mechanics where the concept of force is not primitive. In \cite{Sant'Anna-96} we presented an axiomatic framework for a classical particle mechanics system inspired on Hertz's ideas about mechanics. In the present paper we show an axiomatic framework for classical particle mechanics systems inspired on Mach's ideas about the origin of inertia. As a matter of fact, our main motivation is the work by A. K. T. Assis and P. Graneau \cite{Assis-96}, where they claim that the celebrated experience with Newton's bucket could not be explained, from the physical point of view, by means of a motion with respect to an absolute space. Matter interacts with matter and not with space. Such an idea could be identified with a Mach's principle of inertia.

	Rigorously speaking, there is nothing in the writings of E. Mach which could be called Mach's principle. But since he was one of the most influencial criticists about the absolute newtonian space, and since many authors have suggested that there is some sort of {\em Mach's principle\/} with respect to inertia (for some reviews, see \cite{Assis-89}, \cite{Assis-96} \cite{Assis-99}), we state, in this paper, as Mach's principle the following:

\begin{quote}
The inertia of a material body is due to its interaction with other material bodies.
\end{quote}

	We recognize that such a sentence should be made precise, in the sense that we should present a rigorous definition of {\em inertia\/}, {\em material body\/} and {\em interaction\/}. Nevertheless, we are very confident that the sentence given above may be identified by many physicists as something which we could call Mach's principle. Besides, this principle seems to be physically meaningful in an intuitive level. It seems to be more natural an interaction of matter with matter than matter with (absolute) space, at least into the context of a non-relativistic theory.

	In this paper we make some criticisms concerning the implementation of Mach's principle as proposed in \cite{Assis-96}. Next, with the help of the axiomatic method, we implement some sort of Mach's principle in classical particle mechanics in a mathematically rigorous fashion. We call our axiomatic system as {\em Mach's mechanics\/} or {\em machian mechanics\/}. We show, for example, that contrarily to what is done in \cite{Assis-96}, Mach's principle does not need to be committed with any Weber's like gravitational force. Besides, Mach's mechanics is consistent with any kind of force which obeys the newtonian action-reaction principle, including Newton's gravitational force itself.

	One of the main purposes of this paper is to illustrate the methodological power of the axiomatic method in physics. The axiomatic method is the soul of mathematics, the synthesis of the scientific method, and a powerful language for the scientific philosopher. The logical investigations of physical theories allows us to discuss what can be done and what cannot be done in physics, from the mathematical point of view. For example, can we abandon the usual notions of space and time in physics \cite{Sant'Anna-99}? Can we abandon the notion of force from mechanics \cite{Sant'Anna-96}? Can we incorporate Mach's principle of inertia independently of any gravitational force? In the present paper we answer this last question; and such an answer is positive.

\section{Relational Mechanics}

	In \cite{Assis-96} it is presented a {\em relational mechanics\/} which is supposed to incorporate Mach's principle. According to the authors such a relational mechanics should satisfy the following `axioms':

\begin{enumerate}

\item Force does satisfy the standard newtonian action-reaction principle.

\item The summation of all forces over a material body is always null in all frames.

\end{enumerate}

	It is quite obvious that these sentences are not axioms at all, from the logico-mathematical point of view \cite{Mendelson-97}. Nevertheless they may help us to understand relational mechanics.

	According to \cite{Assis-96}, any implementation of Mach's principle in relational mechanics is not possible if we use the newtonian gravitational force and consider that the distant material bodies, like those outside the Milky Way, may be regarded as an isotropical distribution of matter around any material body $p$ in, e.g., Earth. The resultant force on $p$ due to this isotropical distribution of mass should be zero, according to Newton's gravitation law. So, the second `axiom' above would never be satisfied if $p$ is subject to any other force like, e.g., its gravitational interaction with the planet Earth. One very nice solution to this problem is presented in \cite{Assis-96}. The authors suggest the use of a gravitational force inspired on Weber's electrodynamical interaction \cite{Assis-94}:

\begin{equation}\label{Wforce}
{\bf F}_{12} = -\gamma m_1 m_2 \frac{{\bf r}_{12}}{r_{12}^3}\left( 1 - \frac{6}{c^2}\left( \frac{\dot{r}_{12}^2}{2} - r_{12}\ddot{r}_{12}\right) \right),
\end{equation}

\noindent
where ${\bf F}_{12}$ is the force that particle $2$ exerts on particle $1$, $\gamma$ may be identified with the gravitational constant, $c$ is the speed of light in the vacuum, $m_1$ and $m_2$ are respectively the (gravitational) masses of particles $1$ and $2$, $r_{12} = \parallel {\bf r}_{12}\parallel$ is the norm of the position vector of particle $1$ with respect to particle $2$, and $\dot{r}_{12}$ and $\ddot{r}_{12}$ are respectively the velocity and the acceleration of particle $1$ with respect to particle $2$, which are coincident with the velocity and acceleration of particle $2$ with respect to particle $1$. A most important implication of this Mach-Weber theory is the necessity of a large amount of uniformly distributed matter in the universe. From now on we refer to equation (\ref{Wforce}) as Weber's law of gravitation or Weber's gravitation law.

	Equation (\ref{Wforce}) was inspired on Weber's electrodynamical force:

\begin{equation}\label{Wforce2}
{\bf F}_{12} = \frac{q_1 q_2}{4\pi\varepsilon_0}\frac{{\bf r}_{12}}{r_{12}^3}\left( 1 - \frac{\dot{r}_{12}^2}{2c^2} + \frac{r_{12}\ddot{r}_{12}}{c^2}\right),
\end{equation}

\noindent
where $c = (\varepsilon_0\mu_0)^{-1/2}$, and $q_1$ and $q_2$ are electric charges of particles $1$ and $2$ respectively.

	From equation (\ref{Wforce}) a straight calculation is made in order to derive the gravitational force on a material body (e.g., a particle) $p$ with mass $m$, due to its interaction with the isotropical part of the universe:

\begin{equation}
{\bf F}_{mI} = -\Phi m\left(\ddot{\bf r}_{mS} + {\mbox{\boldmath $\omega$}}_{US}\times ({\mbox{\boldmath $\omega$}}_{US}\times {\bf r}_{mS}) + 2\dot{\bf r}_{mS}\times{\mbox{\boldmath $\omega$}}_{US} + {\bf r}_{mS}\times\frac{d{\mbox{\boldmath $\omega$}}_{US}}{dt}\right),
\end{equation}

\noindent
where

\begin{equation}
\Phi = \frac{2\pi}{3}\xi\frac{\gamma\rho_0}{H_0^2},
\end{equation}

\noindent
$H_0$ is the Hubble's constant, $\pi \approx 3.14159$, $\rho_0$ is the mean density of the universe, ${\bf r}_{mS}$ is the position of particle $p$ with respect to an arbitrary coordinate system $S$, ${\mbox{\boldmath $\omega$}}_{US}$ is the angular velocity of the isotropical distribution of distant galaxies with respect to $S$, and $t$ is the parameter time. The constant $\xi$ is relevant for deriving the precession of the perihelion of the planets. For details see \cite{Assis-96,Assis-99}.

	If $S'$ is a coordinate system where ${\mbox{\boldmath $\omega$}}_{US'}$ is zero then:

\begin{equation}
{\bf F}_{mI} = -\Phi m \ddot{\bf r}_{mS}.
\end{equation}

	If particle $p$ is subject to any force due to an anisotropical distribution of $N$ masses described by

\begin{equation}
{\bf F}_{mA} = \sum_{j=1}^N {\bf F}_{mj},\label{equa6}
\end{equation}

\noindent
then, according to the second `postulate' of relational mechanics,

\begin{equation}
\sum_{j=1}^N {\bf F}_{mj} = m\ddot{\bf r}_{mS},
\end{equation}

\noindent
where $S$ may be any coordinate system which moves with a constant velocity with respect to the system of distant galaxies.

	So, Second Newton's Law is a theorem and inertia of particle $p$ ($\sum_{j=1}^N{\bf F}_{mj}$) is a consequence of its interaction with the distant galaxies.

	If the whole universe does have a rotation ${\mbox{\boldmath $\omega$}}_{US}$ with respect to an arbitrary system of coordinates, then the force that the anisotropical part of the universe exerts on particle $p$ (with mass $m$) is given by:

\begin{equation}
{\bf F}_{mI} = -m\left(\ddot{\bf r}_{mS} + {\mbox{\boldmath $\omega$}}_{US}\times ({\mbox{\boldmath $\omega$}}_{US}\times {\bf r}_{mS}) + 2\dot{\bf r}_{mS}\times{\mbox{\boldmath $\omega$}}_{US} + {\bf r}_{mS}\times\frac{d{\mbox{\boldmath $\omega$}}_{US}}{dt}\right).\label{equa8}
\end{equation}

	This means that a rotating universe is capable of deforming the surface of the water in Newton's bucket, and so Mach's principle is implemented.

	Now, we proceed to some remarks and criticisms on this approach. Next we present solutions to the problems that we point out.

\section{First Remarks and Criticisms}

	We have some remarks and criticisms about the implementation of Mach's principle as proposed in \cite{Assis-96}:

	The first criticism referes to the fact that relational mechanics cannot describe the dynamics of particles without the help of Weber's gravitation force. It is easy to see that the gravitation \`a la Weber, as presented in equation (\ref{Wforce}) is invariant under Galilean transformations if we assume that ${\bf F}_{12}$ is also invariant. At a first sight, this looks like a redundant sentence, but that is not the case. The invariance of  Newton's law of gravitation, for example, is easily proven if we keep in mind that 

\begin{enumerate}

\item force is equal to mass times acceleration (for conservative masses) and

\item force is the gradient of a potential that depends on the diference between space coordinates. 

\end{enumerate}

	In other words, the potential (which depends on a relational concept, namely, the distance between particles) determines the dynamics of particles. In relational mechanics, as proposed in \cite{Assis-96} that does not occur, since Weber's gravitational force itself does not determine any dynamics of any particle. There is no differential equation in relational mechanics analogous to Newton's second law of mechanics. Assis and Graneau claim that Newton's second law demands the notion of an absolute space that interacts with matter, which contradicts Mach's principle. Nevertheless we should recall that Newton's second law does have a very important role, namely, to determine the dynamics of material bodies as a function of the forces acting on them. Newton's second law is a {\em differential equation\/}. We understand the criticisms against Newton's second law. But we do not understand the absence of any differential equation in relational mechanics that determines the dynamics of material bodies. As a matter of fact, there is no relational mechanics at all, since the dynamics of material bodies depends on Weber's law of gravitation and on the use of a machian principle that postulates an isotropical distribution of mass in the universe. Newtonian mechanics allows us to study the dynamics of particles for any gravitational law which obeys the action-reaction principle. In other words, newtonian mechanics is consistent with Newton's gravitation law, Weber's gravitation law and a lot of possibilities that we could imagine and propose. So, there is no relational mechanics, but there is a gravitational theory which postulates Weber's law of gravitation and Mach's principle, but without any underlying mechanics. 

	The second criticism is the fact that in Assis and Graneau theory that Mach's principle cannot be implemented if we use the newtonian force of gravitation.

	These two criticisms may be summarized as it follows: in \cite{Assis-96} there is some sort of confusion between mechanics and gravitation.

	We have a proposal of a relational mechanics system, which we call Mach's mechanics or machian mechanics, that allows us to solve the dynamical problem presented in this section, preserves some of the main ideas by Assis and Graneau and is still consistent with a newtonian law of gravitation. We use the axiomatic method in order to settle our language in a rigorous manner. First we present an axiomatic framework for particle mechanics inspired into newtonian mechanics. In section \ref{Machch} we present an axiomatic framework for particle mechanics inspired on Mach's principle. The axiomatization allows us a comparison between newtonian and machian mechanics.

\section{Further Criticisms: Problems of Symmetry}\label{simetria}

	Discussions on invariance in theoretical physics are rather important. Invariance of physical laws is grounded on an almost `holy' principle usually refered to as the {\em principle of symmetry\/}. This is one of the basis of experimental physics: there must be a manner to establish a communication channel among different observers. We want to focus our discussion on relational mechanics from this point of view.

\begin{enumerate}

\item Webers's gravitation law is an instantaneous action-at-a-distance force which is invariant under Galilean transformations, as we discussed earlier and the reader can easily verify. Nevertheless, no {\em instantaneous\/} action-at-a-distance force can be invariant under Lorentz transformations, since the notion of {\em simultaneity\/} is relative to the observer, according to the special theory of relativity.

\item Weber's electrodynamics is an instantaneous action-at-a-distance force too, which contradicts well known experimental data about accelerating electric charges (like the case of the electrons in an antenna). In this situation, there is no instantaneous interaction. Maxwell's eletromagnetism, which is supported by experience, stands up the existence of eletromagnetic waves that propagate into the space with the speed of light in the vacuum. So, Maxwell's electromagnetism is a more appropriate theory (from the experimental point of view) than Weber's electrodynamics. But Maxwell's physical laws are {\em not\/} invariant under Galilean transformations; they are invariant under Lorentz transformations \cite{Jackson-75}. Besides, it has been shown (i) that Weber's electrodynamical force gives the wrong direction for the radiation field generated by an accelerating charge and (ii) that it cannot explain the synchrotron light (widely confirmed by particle accelerators) since Weber's force takes into account only radial accelerations. For details see \cite{Cavalleri-98a,Cavalleri-98b}. Quoting Cavalleri et al. \cite{Cavalleri-98a}:

\begin{quote}
Concluding, Weber's expression and {\em all\/} its generalizations are wrong without any hope of retrieval.
\end{quote}

\item Hence, it is not possible to use Weber's gravitation law and Maxwell's electrodynamics in the same mathematical framework, i.e., with the same transformation group. If one insists to use Weber's electrodynamics, some well known and basic experimental data should be ignored. Besides, a relational mechanics as suggested by Assis and Graneau, together with a maxwellian eletrodynamics would send us back to the old discussions concerning an ether in space, in order to explain the propagation of electromagnetic waves, by means of the mechanical theory of waves \cite{Einstein-91}.

\end{enumerate}

	One possible solution to this dilema (but we are not sure) is to modify Weber's electrodynamics in order to get a delayed action-at-a-distance theory inspired on the ideas presented by F. Hoyle and J. V. Narlikar in their fabulous book \cite{Hoyle-96}. Gravitation should also be modified into the scope of this proposal.

	Yet, in \cite{Assis-96} Assis and Graneau make a strange comparison between actions-at-a-distance in classical theories and nonlocal phenomena in quantum physics. They claim that `physics has moved imperceptibly away from field-contact action', due to the nonlocal phenomena in the microscopic world. This argument seems to be used in order to make more plausible the acceptance of actions-at-a-distance in classical physics. Some electrodynamical experiments are cited, in order to support their ideas. For a very interesting discussion about some of these experiments see \cite{Cavalleri-98a,Cavalleri-98b}. We do not think that any relationship between classical actions-at-a-distance and quantum nonlocal phenomena are so obviously perceptible. For example, nonlocal phenomena seem to be closely related to the problem of indistinguishability \cite{Mandel-91}, which seems to demand a new mathematical framework \cite{Krause-99,Sant'Anna-00}. That does not seem to be the case in classical physics. For a proposal of a classical picture of nonlocality see \cite{Suppes-96}.

	There is one final remark that we would like to do in this section. It has to do with history. On March 19, 1845, C. F. Gauss wrote a letter to W. Weber \cite{Hoyle-96}:

\begin{quote}
... I would doubtless have published my researches long since were it not that at the time I gave them up I had failed to find what I regarded as the keystone, {\em Nil actum reputans si quid superesset agendum\/}: namely, the derivation of the additional forces -- to be added to the interaction of electrical charges at rest, when they are both in motion -- from an action which is propagated not instantaneously but in time as is the case with light...
\end{quote}

	According to Hoyle and Narlikkar (op. cit.) both Weber and Gauss were worry about a reformulation of action-at-a-distance. Nevertheless, when Weber proposed in 1846 his electrodynamical force given by equation (\ref{Wforce2}) \cite{Assis-94}, he did not intend to solve this problem. He was interested on a general expression for interaction among electric charges such that Coulomb's force and Amp\`ere's force were just particular cases up to appropriate approximations.

\section{McKinsey-Sugar-Suppes System for Classical Particle Mechanics}

	This section is essentially based on the axiomatization of classical particle mechanics due to P. Suppes \cite{Suppes-57}, which is a variant of the formulation by J. C. C. McKinsey, A. C. Sugar and P. Suppes \cite{Suppes-53}. We call this McKinsey-Sugar-Suppes system of classical particle mechanics and abbreviate this terminology as MSS system. MSS system will be useful in order to allow us a comparison between newtonian mechanics (represented by MSS system) and relational mechanics (represented by our Mach's system). In other words, our axiomatic system for Mach's mechanics is achieved thanks to the technique previously used by McKinsey, Sugar and Suppes.

	The reader should not understand that MSS system does faithfully translate all the ideas behind newtonian mechanics; but it translates, in an intuitive manner, some of the main aspects of Newton's ideas. An analogous comment may be done with respect to relational mechanics and our machian system of particle mechanics.

	MSS system has six primitive notions: $P$, $T$, $m$, ${\bf s}$, ${\bf f}$, and ${\bf g}$. $P$ and $T$ are sets, $m$ is a real-valued unary function defined on $P$, ${\bf s}$ and ${\bf g}$ are vector-valued functions defined on the Cartesian product $P\times T$, and ${\bf f}$ is a vector-valued function defined on the Cartesian product $P\times P\times T$. Intuitivelly, $P$ corresponds to the set of particles and $T$ is to be physically interpreted as a set of real numbers measuring elapsed times (in terms of some unit of time, and measured from some origin of time). $m(p)$ is to be interpreted as the numerical value of the mass of $p\in P$. ${\bf s}_{p}(t)$, where $t\in T$, is a $3$-dimensional vector which is to be physically interpreted as the position of particle $p$ at instant $t$. ${\bf f}(p,q,t)$, where $p$, $q\in P$, corresponds to the internal force that particle $q$ exerts over $p$, at instant $t$. And finally, the function ${\bf g}(p,t)$ is to be understood as the external force acting on particle $p$ at instant $t$.

	Now, we can give the axioms for MSS system.

\begin{definicao}
${\cal P} = \langle P,T,{\bf s},m,{\bf f},{\bf g}\rangle$ is a MSS system if and only if the following axioms are satisfied:

\begin{description}
\item [P1] $P$ is a non-empty, finite set.
\item [P2] $T$ is an interval of real numbers.
\item [P3] If $p\in P$ and $t\in T$, then ${\bf s}_{p}(t)$ is a
$3$-dimensional vector (${\bf s}_p(t)\in\Re^3$) such that $\frac{d^{2}{\bf s}_{p}(t)}{dt^{2}}$ exists.
\item [P4] If $p\in P$, then $m(p)$ is a positive real number.
\item [P5] If $p,q\in P$ and $t\in T$, then ${\bf f}(p,q,t) = -{\bf f}(q,p,t)$.
\item [P6] If $p,q\in P$ and $t\in T$, then $[{\bf s}_{p}(t), {\bf f}(p,q,t)] =
-[{\bf s}_{q}(t), {\bf f}(q,p,t)]$.
\item [P7] If $p,q\in P$ and $t\in T$, then
$m(p)\frac{d^{2}{\bf s}_{p}(t)}{dt^{2}} = \sum_{q\in P}{\bf f}(p,q,t) + {\bf g}(p,t).$
\end{description}
\end{definicao}

	The brackets [,] in axiom {\bf P6} denote external product.

	Axiom {\bf P5} corresponds to a weak version of Newton's Third Law: to every force there is always a counterforce. Axioms {\bf P6} and {\bf P5}, correspond to the strong version of Newton's Third Law. Axiom {\bf P6} establishes that the direction of force and counterforce is the direction of the line defined by the coordinates of particles $p$ and $q$.

	Axiom {\bf P7} corresponds to Newton's Second Law.

\begin{definicao}
Let ${\cal P} = \langle P,T,{\bf s},m,{\bf f},{\bf g}\rangle$ be a MSS system, let $P'$ be a non-empty subset of $P$, let ${\bf s}'$, ${\bf g}'$, and $m'$ be, respectively, the restrictions of functions ${\bf s}$, ${\bf g}$, and $m$ with their first arguments restricted to $P'$, and let ${\bf f}'$ be the restriction of ${\bf f}$ with its first two arguments restricted to $P'$. Then ${\cal P'} = \langle P',T,{\bf s}',m',{\bf f}',{\bf g}'\rangle$ is a subsystem of ${\cal P}$ if $\forall p,q\in P'$ and $\forall t\in T$, 
\begin{equation}
m'(p)\frac{d^{2}{\bf s}'_{p}(t)}{dt^{2}} = \sum_{q\in P'}{\bf f}'(p,q,t) + {\bf g}'(p,t).
\end{equation}
\label{P7}
\end{definicao}

\begin{teorema}
Every subsystem of a MSS system is again a MSS system.\footnote{In the original MSS system it is presented another definition for subsystem, where this theorem is not valid.}
\end{teorema}

\begin{definicao}
Two MSS systems \[{\cal P} = \langle P,T,{\bf s},m,{\bf f},{\bf g}\rangle\] and \[{\cal P'} = \langle P',T',{\bf s}',m',{\bf f}',{\bf g}'\rangle\] are equivalent if and only if $P=P'$, $T=T'$, ${\bf s}={\bf s}'$, and $m=m'$.
\end{definicao}

\begin{definicao}
A MSS system is isolated if and only if for every $p\in P$ and $t\in T$, ${\bf g}(p,t) = \langle 0,0,0\rangle$.
\end{definicao}

\begin{teorema}
If \[{\cal P} = \langle P,T,{\bf s},m,{\bf f},{\bf g}\rangle\] and \[{\cal P'} = \langle P',T',{\bf s}',m',{\bf f}',{\bf g}'\rangle\] are two equivalent systems of particle mechanics, then for every $p\in P$ and $t\in T$
\[\sum_{q\in P}{\bf f}(p,q,t) + {\bf g}(p,t) = \sum_{q\in P'}{\bf f}'(p,q,t) + {\bf g}'(p,t).\]\label{somaforcas}
\end{teorema}

	The embedding theorem is the following:

\begin{teorema}
Every MSS system is equivalent to a subsystem of an isolated system of particle mechanics.\label{Her}
\end{teorema}

	The next theorem can easily be proved by Padoa's method:

\begin{teorema}
Mass and internal force are each independent of the remaining primitive notions of MSS system.
\end{teorema}

	According to Suppes \cite{Suppes-57}:

\begin{quote}
Some authors have proposed that we convert the second law [of Newton], that is, {\rm \bf P7}, into a definition of the total force acting on a particle. [...] It prohibits within the axiomatic framework any analysis of the internal and external forces acting on a particle. That is, if all notions of force are eliminated as primitive and {\rm \bf P7} is used as a definition, then the notions of internal and external force are not definable within the given axiomatic framework.
\end{quote}

\section{Set-theoretical Predicate for Mach's Mechanics}\label{Machch}

	As in the previous section, our system has six primitive notions: $P$, $T$, $m$, ${\bf s}$, ${\bf f}$, and ${\bf g}$. $P$ and $T$ are sets, $m$ is a real-valued unary function defined on $P$, ${\bf s}$ and ${\bf g}$ are vector-valued functions defined on the Cartesian product $P\times T$, and ${\bf f}$ is a vector-valued function defined on the Cartesian product $P\times P\times T$. Intuitivelly, $P$ corresponds to the set of particles and $T$ is to be physically interpreted as a set of real numbers measuring elapsed times (in terms of some unit of time, and measured from some origin of time). $m(p)$ is to be interpreted as the numerical value of the mass of $p\in P$. ${\bf s}_{p}(t)$, where $t\in T$, is a $3$-dimensional vector which is to be physically interpreted as the position of particle $p$ at instant $t$. ${\bf f}(p,q,t)$, where $p$, $q\in P$, corresponds to the internal force that particle $q$ exerts over $p$, at instant $t$. And finally, the function ${\bf g}(p,t)$ is to be understood as the external force acting on particle $p$ at instant $t$.

\begin{definicao}
${\cal M} = \langle P,T,{\bf s},m,{\bf f}, {\bf g}\rangle$ is a {\em machian system of particles\/} if and only if the following axioms are satisfied:

\begin{description}
\item [M1] $P$ is a non-empty, finite set.
\item [M2] $T$ is an interval of real numbers.
\item [M3] If $p\in P$ and $t\in T$, then ${\bf s}_{p}(t)$ is a
$3$-dimensional vector (${\bf s}_p(t)\in\Re^3$) such that $\frac{d^{2}{\bf s}_{p}(t)}{dt^{2}}$ exists.
\item [M4] If $p\in P$, then $m(p)$ is a positive real number.
\item [M5] If $p,q\in P$ and $t\in T$, then ${\bf f}(p,q,t) = -{\bf f}(q,p,t)$.
\item [M6] If $p,q\in P$ and $t\in T$, then $[{\bf s}_{p}(t), {\bf f}(p,q,t)] =
-[{\bf s}_{q}(t), {\bf f}(q,p,t)]$.
\item [M7] If $p,q\in P$ and $t\in T$, then

$$\sum_{q\in P}{\bf f}(p,q,t) + {\bf g}(p,t)= {\bf 0}.$$

\item [M8] If $p\in P$ and $t\in T$, then ${\bf g}(p,t) = - m(p)\frac{d^{2}{\bf s}_{p}(t)}{dt^{2}}$.

\end{description}
\end{definicao}

	The brackets [,] in axiom {\bf M6} denote external product, like in the previous section.

	Axioms {\bf M1}-{\bf M6} are in agreement with MSS system. Axioms {\bf M5}-{\bf M6} are in correspondence with the first `postulate' of relational mechanics. Axiom {\bf M7} corresponds to the second `postulate' of relational mechanics and is one of the most important aspects of the proposal presented in \cite{Assis-96}. Notwithstanding, axiom {\bf M8} is a new ingredient, since it is a differential equation which says that the external force ${\bf g}(p,t)$ determines the dynamics of particle $p$. We advocate the idea that this axiom incorporates Mach's principle of inertia if, and only if, we interpret ${\bf g}(p,t)$ as an interaction of $p$ with other `material bodies', whatever is that supposed to mean. In our language there is no necessity to give a detailed description of the nature of this external force. A more detailed and intuitive meaning of ${\bf g}(p,t)$ may be done by means of the models of ${\cal M}$. We discuss this point in the next section. Anyway, axiom {\bf M8} seems to be not in agreement with the relational mechanics proposed by Assis in \cite{Assis-99}. But that does not represent any problem for us, since we are interested on Mach's principle as presented in the Introduction.

	If $p$ is the only particle in $P$, then according to axioms {\bf M7} and {\bf M8} we have the following differential equation:

\begin{equation}
\frac{d^{2}{\bf s}_{p}(t)}{dt^{2}} = {\bf 0},
\end{equation}

\noindent
which corresponds to an `inertial' particle.

	On the other hand, if there is more than one particle $p$ in $P$, then:

\begin{equation}\label{inercia}
\sum_{q\in P}{\bf f}(p,q,t) = m(p)\frac{d^{2}{\bf s}_{p}(t)}{dt^{2}}
\end{equation}

\noindent
which corresponds to Newton's second law.

	In other words, inspired on Mach's ideas about inertia, we proved the existence of inertia in a particle mechanics theory, without any reference to Weber's law of gravitation. As a matter of fact, equation (\ref{inercia}) is consistent with either a newtonian gravitational law, or a Weber gravitation or any other gravitation law which obeys the action-reaction principle given by axioms {\bf M5} and {\bf M6}.

	Assis and Graneau \cite{Assis-96} say that ``gravitational mass is the mass which appears in Newton's law of universal gravitation, in the weight of a body, and in the gravitational potential energy''. If we specify the kind of internal forces $f$ in our system, such forces may (or not) depend on the mass $m$ of particles. Nevertheless, there is no distinction at all between inertial or gravitational mass in our system. We have just {\em mass\/}.

\section{Gravitation in Mach's System}

	In this section we show that our system is consistent with Assis-Graneau proposal with a Weber like law of gravitation as well as with Newton's law of gravitation.

	First we discuss the interpretation of our axiomatic system into the context of Weber's law of gravitation.

	The question is: how to interpret the primitive concepts in a machian system of particles? We show below that Assis-Graneu cosmology may be considered as a possible interpretation of our system. The interpretation of $P$, $T$, and $m$ are obvious. The problem regards the interpretation of ${\bf f}$, ${\bf g}$, and ${\bf s}$.

	According to equations (\ref{equa6}) and (\ref{equa8}) and the second `postulate' of relational mechanics, we have:

\begin{equation}
\sum_{j=1}^N {\bf F}_{mj} - m\left(\ddot{\bf r}_{mS} + {\mbox{\boldmath $\omega$}}_{US}\times ({\mbox{\boldmath $\omega$}}_{US}\times {\bf r}_{mS}) + 2\dot{\bf r}_{mS}\times{\mbox{\boldmath $\omega$}}_{US} + {\bf r}_{mS}\times\frac{d{\mbox{\boldmath $\omega$}}_{US}}{dt}\right) = {\bf 0},\label{hhh}
\end{equation}

\noindent
for the case where the whole universe does have a rotation ${\mbox{\boldmath $\omega$}}_{US}$ with respect to an arbitrary system of coordinates. 

	According to axioms {\bf M7} and {\bf M8}, it seems clear, at least from the logico-mathematical standpoint, that we can interpret the vectors ${\bf f}$, ${\bf g}$, and ${\bf s}$ as it follows:

\begin{equation}
{\bf f}(p,q,t) = {\bf F}_{mj},
\end{equation}

\noindent
where $m$ is the mass of particle $p$ and $j$ is the particle $q$ itself.

\begin{equation}
{\bf g}(p,t) = - m{\bf a},
\end{equation}

\noindent
where ${\bf a}$ is an acceleration given by:

\begin{equation}
{\bf a} = \ddot{\bf r}_{mS} + {\mbox{\boldmath $\omega$}}_{US}\times ({\mbox{\boldmath $\omega$}}_{US}\times {\bf r}_{mS}) + 2\dot{\bf r}_{mS}\times{\mbox{\boldmath $\omega$}}_{US} + {\bf r}_{mS}\times\frac{d{\mbox{\boldmath $\omega$}}_{US}}{dt}.
\end{equation}

	The vector ${\bf s}$ is the solution of the differential equation:

\begin{equation}
\frac{d^{2}{\bf s}_{p}(t)}{dt^{2}} = {\bf a}.\label{iiii}
\end{equation}

	It is important to remark that this last equation (which is consistent with Eq. (\ref{hhh}), gives us a relationship between ${\bf r}$ and ${\bf s}$, which obviously are not the same (${\bf r}\neq {\bf s}$). From the logico-mathematical point of view this is sound, since we are using Assis-Graneau theory as a model for our axiomatic framework. Eq. (\ref{hhh}) is derived from Weber's law of gravitation and the relational mechanics postulates. We interpret the position vector ${\bf s}$ of our system as the solution of the differential equation (\ref{iiii}). This ends our interpretation into the context of Weber's law of gravitation.

	Next, we interpret our axiomatic system into the scope of Newton's law of gravitation.

	The interpretation of newtonian force of gravitation is quite natural. We interpret $P$, $T$, ${\bf s}$, $m$, and ${\bf g}$ in the standard manner, i.e., set of particles, time interval, position vector, rest mass, and external force, respectively. It should be clear to the reader that these are the standard interpretations of $P$, $T$, ${\bf s}$, $m$, and ${\bf g}$ in MSS system, which is a formulation for newtonian classical mechanics. Vector ${\bf f}$ is interpreted as:

\begin{equation}
{\bf f}(p,q,t) = -\gamma m(p)m(q)\frac{{\bf s}_{p}(t) - {\bf s}_{q}(t)}{|{\bf s}_{p}(t) - {\bf s}_{q}(t)|^3}.
\end{equation}

\noindent
which gives, according to axioms {\bf M7} and {\bf M8}, the dynamics of a particle under the influence of the gravitational force given above.

\section{Final Remarks}

	We do not claim that our axiomatic framework for machian mechanics does solve the problems of symmetry pointed out in section \ref{simetria}. It could not be the case, since our axiomatic framework does not make any reference to transformation groups. One manner to deal with this problem is by means of transformation groups that allow us to establish a relationship among different models for our axiomatic framework. The main purposes of our version for machian mechanics are: (i) to show that Mach's principle may be implemented in mechanics without any reference to a Weber's law of gravitation; and (ii) to prove Newton's second law as a theorem with the help of proper axioms.

	A more satisfactory mathematical framework for relational mechanics would be a gauge-theoretical picture by means of fiber bundles. That would be a manner to classify all transformation groups that are consistent with a machian principle for mechanics. That would be an extension of the program presented in \cite{Arnold-89}.

\section{Acknowledgments}

	We acknowledge with thanks the suggestions by Newton C. A. da Costa, Andr\'e K. T. Assis, Giancarlo Cavalleri, and E. Tonni into the context of e-mail messages that we exchanged with them concerning relational mechanics, classical electrodynamics, and the problem of inertia as well as the axiomatic method in physics.


\begin{thebibliography}{99}

\bibitem{Arnold-89} Arnold, V. I., {\em Mathematical Methods of Classical Mechanics\/} (Springer-Verlag, New York, 1989).

\bibitem{Assis-89} Assis, A. K. T., `On Mach's principle' {\em Found. Phys. Lett.\/} {\bf 2} 301-318 (1989).

\bibitem{Assis-94} Assis, A. K. T., {\em Weber's Electrodynamics\/} (Kluwer, Dordrecht, 1994).

\bibitem{Assis-96} Assis, A. K. T. and P. Graneau, `Nonlocal forces of inertia in cosmology' {\em Found. Phys.\/} {\bf 26} 271-283 (1996).

\bibitem{Assis-99} Assis, A. K. T., {\em Relational Mechanics\/} (Apeiron, Montreal, 1999).

\bibitem{Cavalleri-98a} Cavalleri, G., E. Tonni, and G. Spavieri, `On the rigorous force law between current elements and some of its consequences', {\em Hadronic J.\/} {\bf 21} 459-478 (1998).

\bibitem{Cavalleri-98b} Cavalleri, G, G. Bettoni, G. Spavieri, and E. Tonni, `Experimental proof of standard electrodynamics by measuring the self-force on a part of a current loop', {\em Phys. Rev. E\/} {\bf 58} 2505-2517 (1998).

\bibitem{Dugas-55} Dugas, R., {\em A History of Mechanics\/} (Dover, New York, 1955).

\bibitem{Einstein-91} Einstein, A., `On the ether', in {\em The Philosophy of Vacuum\/} edited by S. Saunders and H. R. Brown, 13-20,  (Clarendom Press, Oxford, 1991). Translated by S. W. Saunders and originally published as `\"Uber den \"Ather', {\em Schweizerische naturforschende Gesellschaft, Verhanflungen\/} {\bf 105} 85-93 (1924).

\bibitem{Hertz-94} Hertz, H. R., {\em Die Prinzipien der Mechanik in Neuem Zusammenhange Dargestellt} (Barth, Leipzig, 1894).

\bibitem{Hertz-56} Hertz, H. R., {\em The Principles of Mechanics}, (Dover, New York, 1956).

\bibitem{Hoyle-96} Hoyle, F. and Narlikar, J. V., {\em Lectures on Cosmology
and Action at a Distance Electrodynamics\/} (World Scientific, London, 1996).

\bibitem{Jackson-75} Jackson, J. D., {\em Classical Electrodynamics\/} (John Wiley \& Sons, New York, 1975).

\bibitem{Krause-99} Krause, D., A. S. Sant'Anna, and A. G. Volkov, `Quasi-set theory for bosons and fermions: quantum distributions', {\em Found. Phys. Lett.\/}, {\bf 12} 51-66 (1999).

\bibitem{Mach-74} Mach, E., {\em The Science of Mechanics}, (The Open Court Publishing Co., 1974).

\bibitem{Mandel-91} Mandel, L., `Coherence and indistinguishability', {\em Optics Letters} {\bf 16} 1882-1884 (1991). 

\bibitem{Mendelson-97} Mendelson, E., {\em Introduction to Mathematical Logic\/} (Chapman \& Hall, London, 1997).

\bibitem{Suppes-53} McKinsey, J. C. C., A. C. Sugar and P. Suppes, 
`Axiomatic foundations of classical particle mechanics', {\em J. Rational Mechanics and Analysis\/}, {\bf 2} 253-272 (1953).

\bibitem{Sant'Anna-96} Sant'Anna, A. S., `An axiomatic framework for classical particle mechanics without force', {\it Philos. Natur.} {\bf 33} 187-203 (1996).

\bibitem{Sant'Anna-99} Sant'Anna, A. S., `An axiomatic framework for classical particle mechanics without space-time', {\em Philos. Natur.\/} {\bf 36} 307-319 (1999).

\bibitem{Sant'Anna-00} Sant'Anna, A. S. and A. M. S. Santos, `Quasi-set-theoretical foundations of statistical mechanics: a research program', {\em Found. Phys.\/}, {\bf 30} 101-120 (2000).

\bibitem{Suppes-57} Suppes, P., {\em Introduction to Logic\/}, (Van Nostrand, Princeton, 1957).

\bibitem{Suppes-96} Suppes, P., J. A. de Barros, and A. S. Sant'Anna, `Violation of Bell's inequalities with a local theory of photons', {\em Found. Phys. Lett.} {\bf 9} 551-560 (1996).

\end{thebibliography}
\end{document}